# Control of the chaotic state caused by the current-driven ion acoustic instability and dynamical behavior using delayed feedback

Takao Fukuyama[a]*, Christian Wilke[a], Yoshinobu Kawai[b]

[a] *Institut fuer Physik, Ernst-Moritz-Arndt-Universitaet Greifswald*
*Domstrasse 10a, D-17489 Greifswald, Germany*
[a] *Guests of Teilinstitut Greifswald, Max-Planck-Institut fuer Plasmaphysik,*
*Wendelsteinstrasse 1, D-17491 Greifswald, Germany*
[b] *Interdisciplinary Graduate School of Engineering Sciences,*
*Kyushu University, Kasuga-kouen 6-1, Kasuga, Fukuoka 816-8580, Japan*
*Email: fukuyama@ipp.mpg.de

**Abstract**

Controlling chaos caused by the current-driven ion acoustic instability is attempted using the delayed continuous feedback method, i.e., the time-delay auto synchronization (TDAS) method introduced by Pyragas [Phys. Lett. A **170** (1992) 421.]. When the control is applied to the typical chaotic state, chaotic orbit changes to periodic one, maintaining the instability. The chaotic state is well controlled using the TDAS method. It is found that the control is achieved when a delay time is chosen near the unstable periodic orbit corresponding to the fundamental mode. Furthermore, when the delayed feedback is applied to a periodic nonlinear regime and arbitrary time delay is chosen, the periodic state is leaded to various motions including chaos. As a related topic, the synchronization between two instabilities of autonomous discharge tubes in a glow discharge is studied. Two tubes are settled independently and interacting each other through the coupler consisted of variable resister and capacitor. When the value of resister is changed as the strength of coupling, coupled system shows a state such as chaos synchronization.

## 1. Introduction

Over the past decade, the problem of controlling chaos has attracted great interest in many fields, such as lasers, motivated by the importance of the role. The serious role of turbulence in fusion-oriented plasmas creates a special interest in controlling chaos. An effective method of controlling chaos, which has been proposed by Ott, Grebogi, and Yorke (OGY),[1] has attracted much attention. On the other hand, Pyragas[2] has proposed a time-delay feedback technique, i.e., the time-delay auto synchronization (TDAS) method,[3] which is appropriate for the experimental systems working in real time. Chaos control is attempted using the TDAS method based on the Pyragas technique in order to attain stable chaos control.

## 2. Experimental setup

The experiments are performed using a Double Plasma device. Argon gas is introduced into the chamber at a pressure of $4.0 \times 10^{-4}$ Torr. Typical plasma parameters are as follows: the electron density $n_e \sim 10^8$ cm$^{-3}$, electron temperature $T_e \sim 0.5$-$1.0$ eV. The current-driven ion acoustic instability is excited by the two parallel mesh grids installed into the chamber ($G_1$ and $G_2$). A dc potential $V_m$ is applied to $G_1$ in order to excite the instability, and $G_2$ is kept at floating potential. Time series signals for analysis are obtained from the fluctuating components of the currents on the $V_m$ biased mesh grids. The experiments in controlling chaos are performed by applying the feedback signal $F(x)$ to the floating mesh grid $G_2$. The feedback signal $F(x)$ is generated from $x(t)$, using the electronic circuit based on the TDAS method. Chart of experimental setting is shown in Fig. 1.

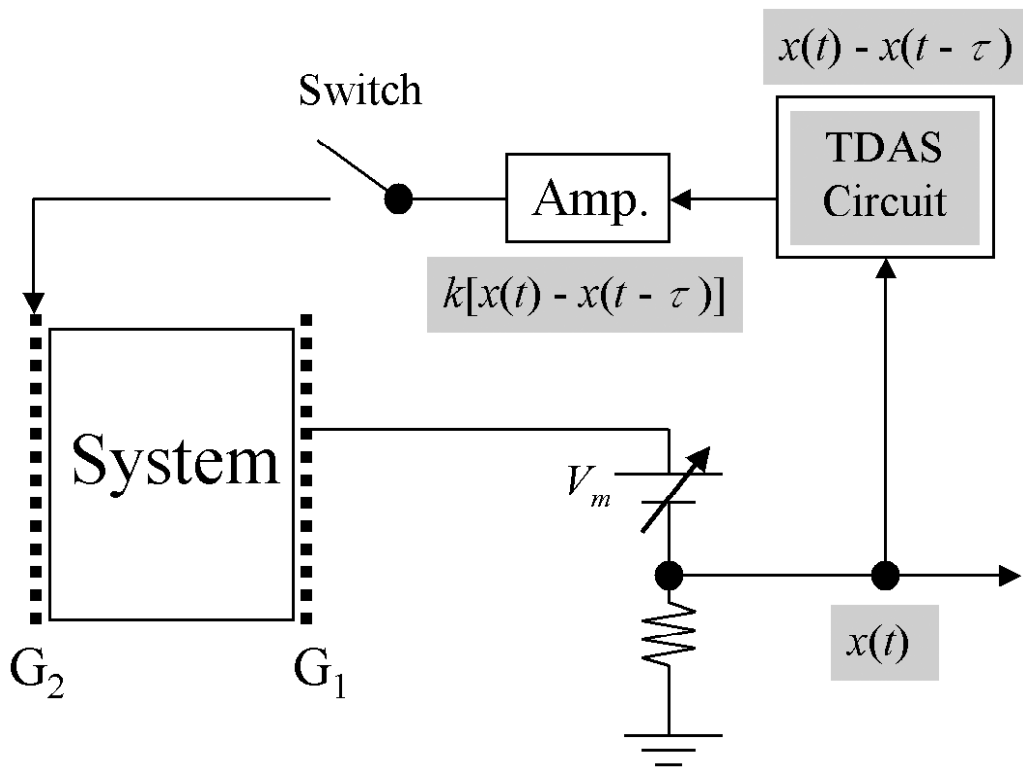

Figure1: Chart of experimental setting.

## 3. Results and discussions

When the grid bias $V_m$ exceeds a threshold, the current-driven ion acoustic instability is excited. When $V_m$ exceeds 40 V, a limit cycle oscillation appears. Then, according to increasing $V_m$, the system becomes chaotic via bifurcation. The system presents a typical chaotic feature around $V_m$ = 54 V. The TDAS control is applied to the typical chaotic state. Figure 2 shows the stabilization process of chaos using the TDAS method. Figure 2(a) and (b) correspond to the time series signal and the feedback signal during the transition from the uncontrolled to the controlled state, respectively. Here, $t$ and $k$ are 20 micro s (~ 1.16 period) and 0.28, respectively. It is found that the system changes from chaotic to periodic, maintaining the instability, and the fundamental mode of the unstable periodic orbit is selected during controlling process.[4]

Furthermore, delayed feedback is applied to periodic nonlinear regime ($V_m$ = 40 V), and the dynamical behavior is studied. When the TDAS method is applied to periodic nonlinear regime and arbitrary delay time $t$ is chosen, the periodic state changes to various states such as intermittency and unstable "period-3" orbits corresponding to period-doubling bifurcation, as shown in Fig. 3.

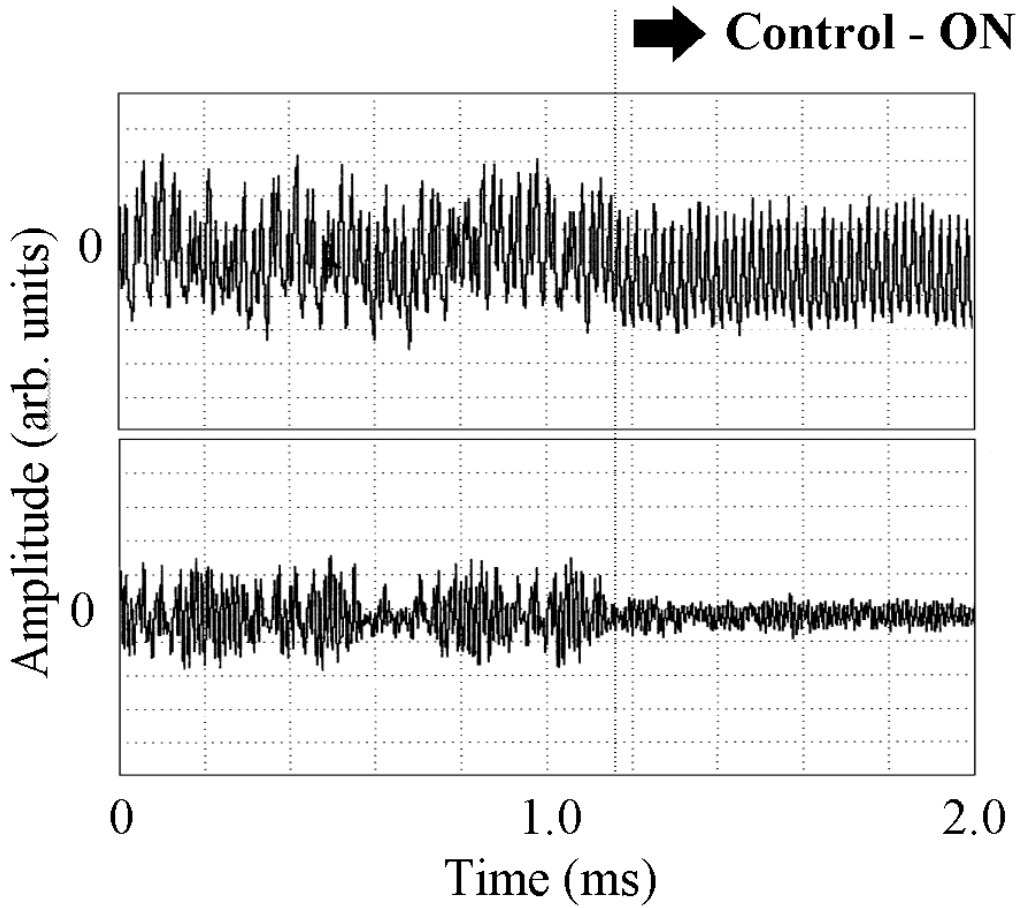

Figure2: The stabilization process of chaos using the TDAS method.

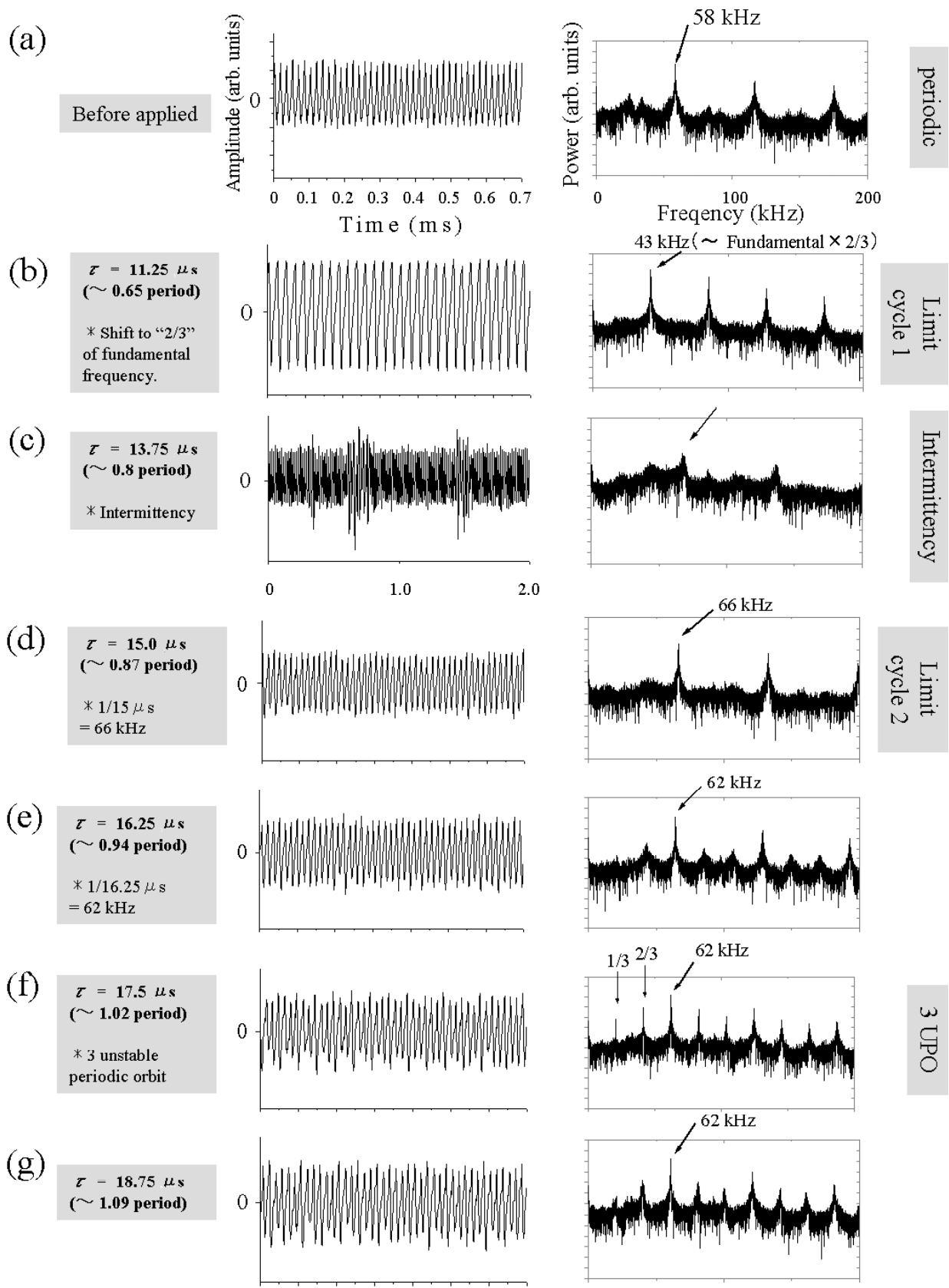

Figure 3: Dynamical behavior of the system as a function of $t$.

## 4. Related topic (The dynamical behavior of two coupled chaotic oscillators)

As a related topic, the synchronization between two instabilities of autonomous discharge tubes in a glow discharge is studied. Recently, the synchronization of two chaotic oscillators [5][6] has attracted much attention in many branches of science, motivated by the possibility of wide spread applications of coupled nonlinear oscillators. It is well known that two chaotic oscillators can synchronize through interaction, namely, coupling. This synchronization has useful applications in chaos control. The behaviors of coupled nonlinear oscillators are interesting phenomena in the study of plasma physics as well as other branches of science.

Two tubes are settled independently and interacting each other through the coupling of variable resister and capacitor, as shown in Fig. 4. Here, when the value of resister is changed as the strength of coupling, coupled systems show a state such as chaos synchronization. Parameters are as follow: pressure of every tube is 4.78 mb, discharge current of tube 1 and 2 are 24.16 mA and 24.25 mA, respectively. Here, values of resister and capacitor, which are settled at a coupler, are 60 k Ohm and 4 micro F. Figure 5(a) and (b) show time series and *X-Y* plot when two oscillators synchronize, respectively.

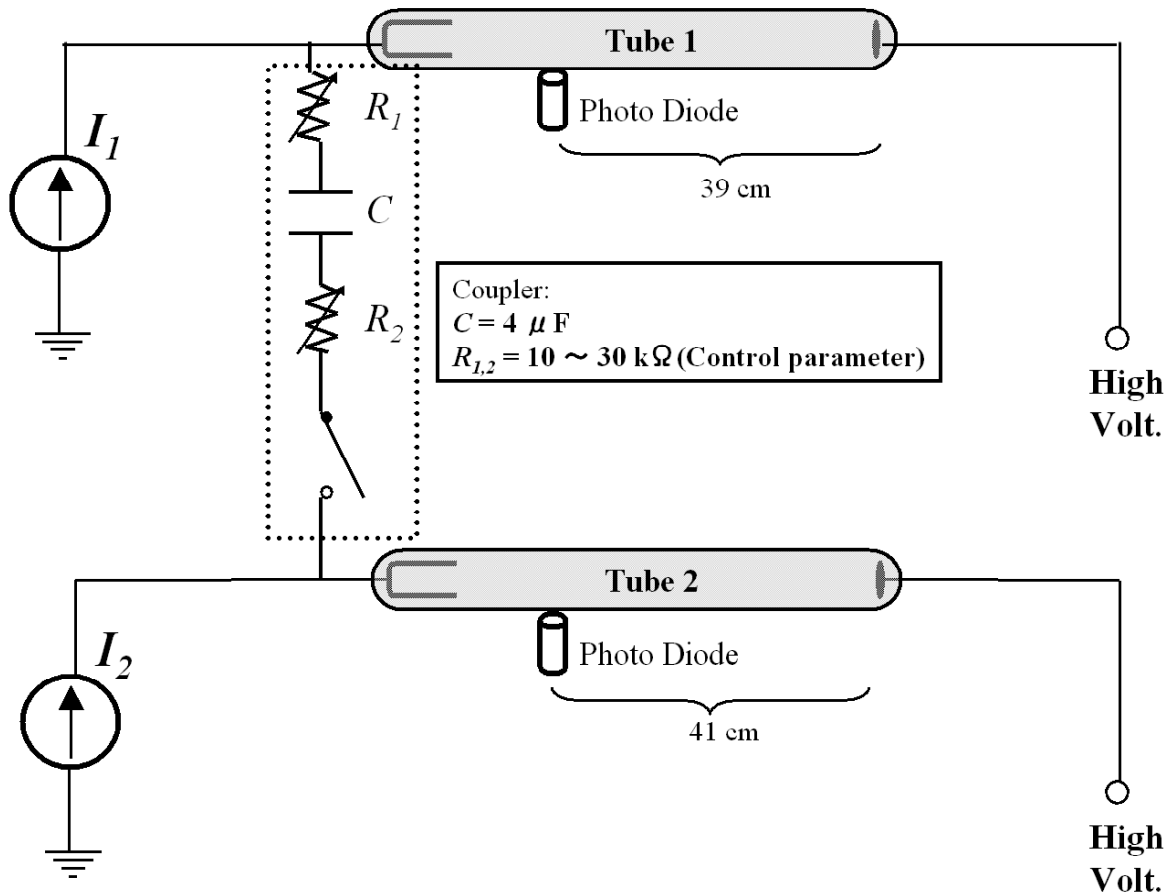

Figure4: Chart of experimental setting of two coupled oscillators.

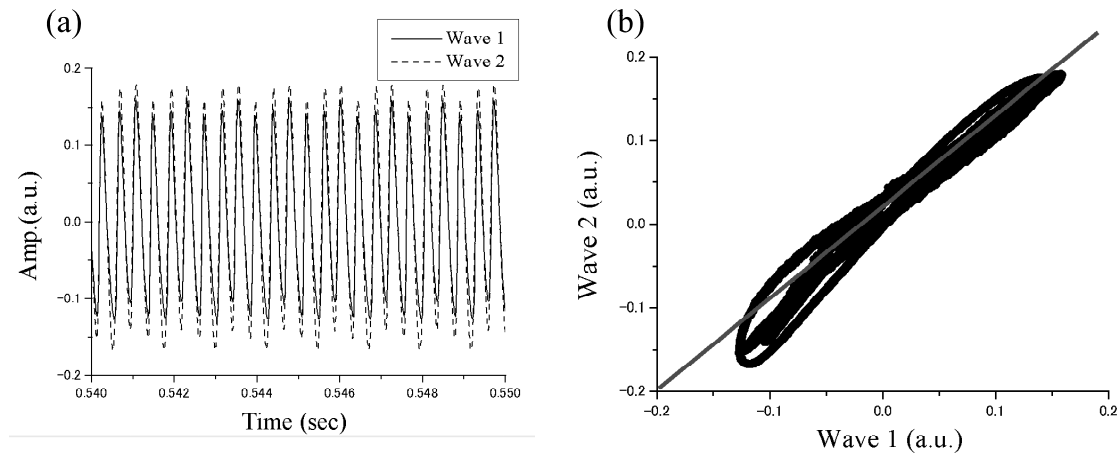

Figure 5: (a) Time series and (b) *X-Y* plot in synchronization are shown.

As a next step, a spatio-temporal structure has been attracted much attention. Up to now, only temporal appearance has been analyzed. Ionization wave system is appropriate for investigations of spatio-temporal structure, since it is easy to measure a structure of the system by using CCD camera. Then, the result observed over space and time, namely "spatio-temporal chaos", is shown in Fig.6. It is observed that the coupled oscillators synchronize over not only time but also "space".

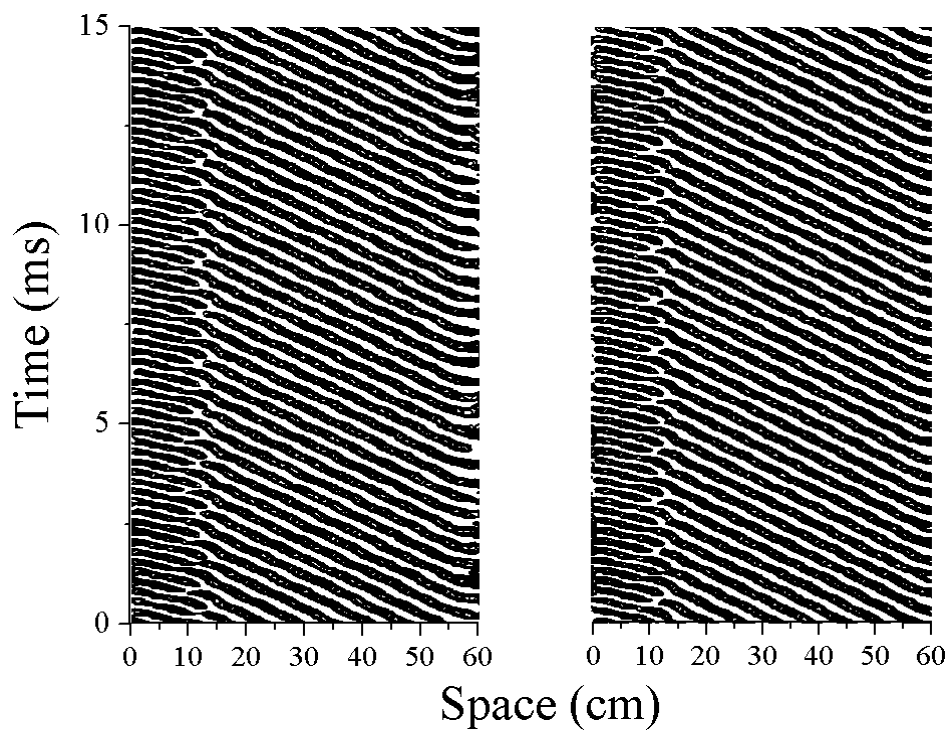

Figure6: The appearance over space and time, namely the synchronized "spatio-temporal chaos" in coupled oscillators, is shown.

## 5. Conclusion

When the TDAS control is applied to the typical chaotic state, chaotic orbit changes to periodic one, maintaining the instability. The chaotic state caused by the current-driven ion acoustic instability is well controlled using the TDAS method. It is found that the control is achieved when a time delay is chosen near the unstable periodic orbit corresponding to the fundamental mode embedding in the chaotic system. Furthermore, when the delayed feedback is applied to a periodic nonlinear regime and arbitrary time delay is chosen, the periodic state is leaded to various motions including chaos.

As a related topic, the synchronization between two instabilities of autonomous discharge tubes in a glow discharge is studied. When the value of resister is changed as the strength of coupling, coupled oscillators show a state such as spatio-temporal chaos synchronization.

**Acknowledgements**

The authors would like to thank to Dr. Ruslan Kozakov, Mr. Holger Testrich, and Mr. Christian Brandt for their fruitful discussions and kind help. Takao Fukuyama is supported by Japan Society for the Promotion of Science, Postdoctral Fellowships for Research Abroad.